\documentstyle[11pt]{article}
\begin{document}
\begin{titlepage}
\begin{flushright}
SUSSEX-AST 93/7-2 \\
astro-ph/9307020\\
(July 1993)\\
\end{flushright}
\begin{center}
\Large
{\bf The Inflationary Energy Scale}\\
\vspace{.3in}
\normalsize
\large{Andrew R. Liddle} \\
\normalsize
\vspace{.6 cm}
{\em Astronomy Centre, \\ School of Mathematical and Physical Sciences,\\
University of Sussex, \\ Brighton BN1 9QH, U.~K.}\\
\vspace{.6 cm}
\end{center}
\baselineskip=24pt
\begin{abstract}
\noindent
The energy scale of inflation is of much interest, as it suggests the scale of
grand unified physics and also governs whether cosmological events such as
topological defect formation can occur after inflation. The COBE results are
used to limit the energy scale of inflation at around 60 $e$-foldings from the
end of inflation. An exact dynamical treatment based on the Hamilton-Jacobi
equations is then used to translate this into limits on the energy scale at
the end of inflation. General constraints are given, and then tighter
constraints based on physically motivated assumptions regarding the allowed
forms of density perturbation and gravitational wave spectra. These are also
compared with the values of familiar models.
\end{abstract}

\begin{center}
\vspace{1cm}
PACS numbers~~~98.80.Cq, 98.70.Vc\\
\vspace{1cm}
Submitted to {\em Physical Review D}
\end{center}

\end{titlepage}

\section{Introduction}

Limits on the energy density at which cosmological inflation \cite{INFL} takes
place are of great interest, being a prime example of a situation where
cosmological observations might provide information regarding the correct
physics at energies completely inaccessible by terrestrial means. An accurate
estimate of the inflationary energy scale may provide vital information
concerning the scale of unification for gauge interactions, for example. The
energy scale is also of interest for cosmological reasons; for instance, one
is interested to know whether or not the inflationary energy scale is so low
as to forbid the formation after inflation of topological defects that might
be of interest for structure formation \cite{HP}. The inflationary scale also
determines whether or not topological defects can quantum mechanically
nucleate during inflation \cite{GV}.

The usual goal of studies such as this is to provide upper limits to the
inflationary energy scale. As we shall see, lower limits are much harder to
come by. Studies made in the eighties and early nineties \cite{INFENG,LEPS}
typically made some simple modelling of inflation, and then imposed what was
at that time the current upper limits on microwave fluctuations. (If further
assumptions such as the existence of an axion field were made, then other
constraints could be brought to bear \cite{LAX}.) Conceptually simpler but in
general weaker limits can be obtained by considering only the effect of
gravitational wave modes \cite{GRAV,WHITE}. The measurement of large scale
microwave fluctuations by the COBE DMR instrument \cite{COBE} now allows one
to be much more definite, in any given model replacing an upper limit with a
definite value (and uncertainty). The understanding of the influence of
gravitational wave modes excited during inflation on the microwave background
\cite{GRAV} has also advanced considerably recently \cite{TENSOR,LL1,LMM}. The
detection of such modes would provide vital information as regards bounding
the energy scale from below, as we shall discuss.

The discussion here is focussed on inflationary models which utilize a single
rolling scalar field; that is, chaotic inflation \cite{LIN} in its loosest
sense. In such models it is usual to assume that one can choose the potential
of this field as one likes. The results derived here are rigidly true only in
this case. However, they also hold in models with multiple rolling scalar
fields, provided that the fluctuations in field directions orthogonal to the
classical trajectory are small; indeed, as these fluctuations would inevitably
reduce the allowed energy scale by soaking up some of the COBE anisotropy, any
upper bounds derived here continue to be true in this case. They do not
directly apply to models which rely on scalar fields trapped in metastable
vacuum states, though even there one can usually, as with extended inflation
\cite{EI}, rephrase this situation in terms of a rolling field.

There are typically two steps in finding the inflationary energy scale. The
first is to limit the energy density at the time when fluctuations observable
in the microwave background were generated. This occurs as those scales left
the horizon during inflation, typically when the scale factor was around
$e^{-60}$ of its size at the end of inflation (normally referred to as 60
$e$-foldings from the end of inflation). Here we aim to provide a much more
general treatment than before, utilizing results to first-order in the
slow-roll approximation, thus incorporating both the predicted tilt
\cite{TILT} in the density perturbation (scalar) spectrum from inflation and
also including the effect of gravitational wave (tensor) modes with their
characteristic scale dependence. This enables accurate bounds to be placed on
the energy density 60 $e$-foldings from the end of inflation.

The second aspect of finding the inflationary energy scale is to use the limit
60 $e$-foldings from the end, and evolve the system so as to provide a limit
on the energy scale at the end of inflation. In the past this has been
accomplished by using the slow-roll approximation; however, inflation can only
end if the slow-roll approximation breaks down, and so such approaches are
necessarily inaccurate. In this paper, the equations are written in
Hamilton-Jacobi form \cite{HJ}, which allows the inflationary dynamics to be
treated {\em exactly}. This involves treating the Hubble parameter, which
directly measures the energy density, as a function of the inflaton field
$\phi$.

The outline of the paper is as follows. In section 2, the equations are set up
in Hamilton-Jacobi form. A new proposal is then implemented for the
specification of inflationary models, where rather than specifying them by a
potential $V(\phi)$, they are instead specified by a function $\epsilon(\phi)$
which measures how accurately the slow-roll approximation holds as a function
of scalar field value. It is emphasized that this classification covers all
inflationary models involving rolling fields, and that the dynamics are
treated exactly, {\em not} subject to any form of slow-roll approximation.
Section 3 discusses the generation of perturbation spectra by a given
inflationary model, and uses this to bound from above the inflationary energy
scale 60 $e$-foldings from the end of inflation. Section 4 takes advantage of
the Hamilton-Jacobi formalism to produce limits on the energy scale at the end
of inflation, both under very general circumstances and more restrictively by
imposing physically motivated constraints on the form of perturbation spectra
produced. Section 5 provides the conclusions, and also discusses the
possibility of producing lower bounds on the energy scale.

\section{Inflationary Dynamics}

The Hamilton-Jacobi equations arise when one rewrites the equations of motion
in a way that allows one to write the Hubble parameter as a function of the
scalar field $\phi$. The usual equations of motion are
\begin{eqnarray}
\label{eq1}
H^2 & = & \frac{8 \pi}{3 m_{Pl}^2} \left( \frac{1}{2} \dot{\phi}^2 + V(\phi)
	\right)\,, \\
\label{eq2}
\ddot{\phi} & + & 3 H \dot{\phi} + V' = 0 \,,
\end{eqnarray}
with $H=\dot{a}/a$ the Hubble parameter, $a$ the scale factor, $m_{Pl}$ the
Planck mass, and where as usual dots are time derivatives and primes
derivatives with respect to the scalar field $\phi$. Differentiating the first
with respect to $t$ and using the second gives
\begin{equation}
2 \dot{H} = - \frac{8\pi}{m_{Pl}^2} \dot{\phi}^2 \,.
\end{equation}
We assume that $\dot{\phi}$ never passes through zero during inflation,
allowing us to use $\phi$ as a time variable\footnote{In rolling models, this
is always a good assumption. It can only be violated while inflation is still
occurring if the potential has a local minimum with non-zero potential energy,
in which case the field will become a trapped one.}. We may therefore divide
each side by it and eliminate the time dependence in the Friedmann equation,
obtaining the Hamilton-Jacobi equations \cite{HJ}
\begin{eqnarray}
\label{e1}
(H')^2 - \frac{12 \pi}{m_{Pl}^2} H^2 & = & - \frac{32\pi^2}{m_{Pl}^4}
	V(\phi) \, ,\\
\label{e2}
\dot{\phi} & = & - \frac{m_{Pl}^2}{4\pi} H' \,.
\end{eqnarray}
We shall throughout make the choice that $\dot{\phi} > 0$. With the equations
in this form, it is natural to think of specifying inflationary models by a
choice of $H(\phi)$ rather than $V(\phi)$ \cite{LID}; one can then easily
generate a large set of exact inflationary solutions simply by
differentiation, whereas a choice of $V(\phi)$ requires the normally
impossible task of analytically solving the coupled differential equations.

We can now define what we shall refer to as slow-roll parameters,
$\epsilon(\phi)$ and $\eta(\phi)$, by \cite{LL1}
\begin{eqnarray}
\label{H1}
\epsilon(\phi) & = & \frac{m_{Pl}^2}{4\pi} \left( \frac{H'}{H} \right)^2 \,,\\
\label{H2}
\eta(\phi) & = & \frac{m_{Pl}^2}{4\pi} \frac{H''}{H} = \epsilon(\phi) -
	\sqrt{\frac{m_{Pl}^2}{16\pi}} \,
	\frac{\epsilon'(\phi)}{\sqrt{\epsilon(\phi)}}\,.
\end{eqnarray}
The sign of the last term, like the signs of other equations featuring
$\sqrt{\epsilon(\phi)}$ later, is determined from the choice $\dot{\phi} > 0$.
Wherever square roots are utilized, it is the positive root that is to be
taken, with the overall sign incorporated in the prefactor. These parameters
measure how accurate the slow-roll approximation would at a given value of
$\phi$; their smallness corresponds, respectively, to the validity of
neglecting the first term in Eq.~(\ref{e1}) and the first term of its
$\phi$-derivative. Let us emphasize again though that we will not make a
slow-roll approximation in considering the dynamics. Further, $\epsilon(\phi)$
possesses the extremely useful property that the condition for inflation,
$\ddot{a} > 0$, is precisely equivalent to $\epsilon(\phi) < 1$.

In this paper, it is convenient to go one small step further than specifying
models by $H(\phi)$; instead we shall specify models by choosing
$\epsilon(\phi)$. By allowing arbitrary forms of this function, we can
specify arbitrary inflationary models just as well as if we were to use
$V(\phi)$. Our choice though allows analytic progress without slow-roll
approximation.

The number of $e$-foldings $N$ between scalar field values $\phi$ and
$\phi_{{\rm end}}$ (the latter being the scalar field value when inflation
ends) is given by
\begin{equation}
N(\phi,\phi_{{\rm end}}) \equiv \ln \frac{a(\phi_{{\rm end}})}{a(\phi)}
	= \sqrt{\frac{4\pi}{m_{Pl}^2}} \int_{\phi}^{\phi_{{\rm end}}}
	\frac{1}{\sqrt{\epsilon(\phi)}} \, {\rm d}\phi \,.
\end{equation}
Unlike the similar equation often seen featuring $V/V'$, this expression is
exact. The end of inflation, when the scale factor stops accelerating, is
given precisely by $\epsilon(\phi)=1$, which determines $\phi_{{\rm
end}}$.\footnote{This statement is true for all rolling models. If one is
considering models which end inflation by an unusual means such as bubble
nucleation in a field other than the rolling one ({\it eg} extended
inflation \cite{EI}), this provides an exception and $\phi_{{\rm end}}$ must
be determined via the physics of the nucleation process. In such cases
$\epsilon$ may be less than unity at the end of inflation, though one could
imagine that it had increased extremely rapidly to unity.}

One computes $H(\phi)$ by quadrature from
\begin{equation}
\frac{{\rm d} \ln H}{{\rm d}\phi} = -
	\sqrt{\frac{4\pi\epsilon(\phi)}{m_{Pl}^2}} \,,
\end{equation}
to get
\begin{equation}
H(\phi) = H_{{\rm end}} \exp \left( \int_{\phi}^{\phi_{{\rm end}}}
	\sqrt{\frac{4\pi\epsilon(\phi)}{m_{Pl}^2}} {\rm d} \phi \right) \,,
\end{equation}
where $H_{{\rm end}}$ is of course $H(\phi_{{\rm end}})$, the Hubble parameter
at the end of inflation. The Hubble parameter is a direct measure of the
energy scale, and so bounding the energy scale simply amounts to bounding $H$
at different epochs.

The potential which generates the solutions is then
\begin{equation}
V(\phi) = \frac{3 m_{Pl}^2}{8\pi} H^2(\phi) \left( 1 -
	\frac{\epsilon(\phi)}{3} \right) \,.
\end{equation}
Whenever slow-roll is good (small $\epsilon$) one has $V(\phi) \propto
H^2(\phi)$. One can thus generate an endless set of exact solutions from
choices of $H(\phi)$, or from $\epsilon(\phi)$ in those cases where the
integration giving $H(\phi)$ can be done analytically.

One can use these equations to calculate the density perturbation amplitude
$\delta_H(\phi)$, as formally defined in \cite{LL2}, which to lowest order in
slow-roll is
\begin{eqnarray}
\delta_H(\phi) & \equiv & \left. \frac{H(\phi)^2}{5\pi |\dot{\phi}|}
	\right|_{aH=k} \;,\\
 & = & \left. \frac{2}{5\sqrt{\pi}} \frac{H(\phi)}{m_{Pl}
	\sqrt{\epsilon(\phi)}} \right|_{aH=k} \;.
\end{eqnarray}
One can then satisfy the COBE result \cite{COBE}, most conveniently taken to
be evaluated 60 $e$-foldings from the end of inflation. We shall henceforth
take $\delta_H$ to indicate the amplitude of the spectrum at this time. This
fixes $H_{{\rm end}}$, provided one knows how to incorporate tilt and
gravitational wave corrections into the correct normalization of $\delta_H$.

Provided inflation ends at $\epsilon(\phi)=1$, one then has
\begin{equation}
V_{{\rm end}} = \frac{m_{Pl}^2}{4\pi} H_{{\rm end}}^2 \,,
\end{equation}
though to estimate the energy density one should include the kinetic
contribution, writing
\begin{equation}
\rho_{{\rm end}} = \frac{3 m_{Pl}^2}{8\pi} H_{{\rm end}}^2 \,.
\end{equation}

It is best to illustrate this formalism via an example, which corresponds
rather closely to the usual polynomial chaotic inflation models \cite{LIN}
with potentials $V(\phi) \propto \phi^{\alpha}$. Let us choose
\begin{equation}
\epsilon(\phi) = \frac{m_{Pl}^2 \alpha^2}{16\pi \phi^2}
\end{equation}
with negative $\phi$ and $\alpha$ a constant. Inflation ends at
$\epsilon(\phi) = 1$, giving $\phi_{{\rm end}}^2 = \alpha^2 m_{Pl}^2/16\pi$,
and we have
\begin{equation}
N(\phi,\phi_{{\rm end}}) = \frac{4\pi}{\alpha} \frac{\phi^2}{m_{Pl}^2} -
	\frac{\alpha}{4} \,.
\end{equation}
Solving, we get
\begin{equation}
H(\phi) = H_{{\rm end}} \left( \frac{\phi}{\phi_{{\rm end}}}
	\right)^{\alpha/2} \,,
\end{equation}
and so
\begin{equation}
H_{60} = H_{{\rm end}} \left( 1 + \frac{240}{\alpha} \right)^{\alpha/4} \,,
\end{equation}
where $H_{60}$ is the Hubble parameter 60 $e$-foldings from the end of
inflation. Thus
\begin{equation}
\delta_H = \frac{2}{5\sqrt{\pi}} \frac{H_{{\rm end}}}{m_{Pl}
\sqrt{\epsilon_{60}}} \, \left( 1 + \frac{240}{\alpha} \right)^{\alpha/4}
	\,,
\end{equation}
with
\begin{equation}
\epsilon_{60} = \frac{\alpha}{240+\alpha} \quad ; \quad \eta_{60} =
	\frac{\alpha-2}{240+\alpha} \,.
\end{equation}
In fact $\alpha = 2$ corresponds to the special case where $\eta(\phi)$ is
identically zero for all $\phi$.

In the following section, we shall see that for models with small
$\epsilon_{60}$ and $\eta_{60}$ such as these, the appropriate $\delta_H$ to
explain the COBE result is $1.7 \times 10^{-5}$ \cite{LL2}. Consequently, one
has
\begin{eqnarray}
\frac{H_{{\rm end}}}{m_{Pl}} & = & 7.5 \times 10^{-5} \left(
	1+\frac{240}{\alpha} \right)^{-\frac{2+\alpha}{4}} \,,\\
 & = &	\left\{	\begin{array}{ll}
		6.2 \times 10^{-7} \quad {\rm for} \; \; \alpha = 2\\
		1.6 \times 10^{-7} \quad {\rm for} \; \; \alpha = 4
		\end{array}
	\right.
\end{eqnarray}

The potential supplying this $\epsilon(\phi)$ is
\begin{equation}
V(\phi) = \frac{3 m_{Pl}^2}{8\pi} H_{{\rm end}}^2 \left( 1 - \frac{m_{Pl}^2
	\alpha^2}{48 \pi \phi^2} \right) \left(
	\frac{\phi}{\phi_{{\rm end}}}\right)^\alpha \,,
\end{equation}
confirming that in the slow-roll limit we just get the polynomial potentials
of the simplest chaotic inflation models. The analytic solution requires that
the potential has the extra $\phi$-dependent correction term which makes the
solutions exact. A suitable adjustment of the original $\epsilon(\phi)$ can be
used to give exactly $V(\phi) \propto \phi^{\alpha}$, though it cannot be
written analytically.

\section{The Perturbation Spectra, and Limiting $H_{60}$}

Returning to the general case, we now need to examine in detail what the COBE
normalization means. In the last section, we mentioned the fiducial
normalization $\delta_H = 1.7 \times 10^{-5}$ which is correct only for
sufficiently flat scalar spectra with negligible gravitational waves. This is
appropriate only if the slow-roll parameters $\epsilon(\phi)$ and $\eta(\phi)$
are small at the time the relevant scales leave the horizon. By utilizing
standard results \cite{LL2}, we can improve this to incorporate the first
level of slow-roll corrections, a treatment which is adequate for all models
which appear viable when confronted with the full range of large scale
structure observations \cite{LL2}.

In the spirit of the above, we shall assume that the scales corresponding to
quadrupole anisotropies passed out of the horizon 60 $e$-foldings from the end
of inflation\footnote{Changing this number (which depends weakly on the
physics of reheating) does not have any significant effect, as we shall see
in the next section.}, and that across the scales which contribute
significantly to the COBE observation (which are only a few $e$-foldings) the
spectral indices of the scalar and tensor modes can be treated as
scale-independent (that is, the spectra are approximated by power-laws). It is
then easy to show \cite{LL1} that the spectral indices are given from the
slow-roll parameters at that time as\footnote{These are correct to first-order
in the slow-roll parameters. Stewart and Lyth \cite{SL} have provided
expressions correct to second-order, these corrections normally being small.
We shall not utilize these here. Note that the numerical factors are different
from those in \cite{LL1}, due to a slightly different definition of the
slow-roll parameters.}
\begin{eqnarray}
n^S_{60} & = & 1- 4\epsilon_{60} + 2 \eta_{60} \,\\
 & = & 1 - 2\epsilon_{60} + \sqrt{\frac{m_{Pl}^2}{4\pi}} \,
	\frac{\epsilon_{60}'}{\sqrt{\epsilon_{60}}}\;,\\
n^T_{60} & = & - 2 \epsilon_{60} \,.\\
\end{eqnarray}

In addition to this, we need to know the contributions of the scalars and
tensors to the microwave anisotropies. As usual, the fractional temperature
anisotropy is split into multipoles (with the monopole and dipole removed)
\begin{equation}
\frac{\Delta T}{T} (\theta,\phi) = \sum_{l,m} a_{lm} Y^l_m(\theta,\phi) \,.
\end{equation}
Inflation predicts the (rotationally invariant) expectation of these
multipoles, $\Sigma_l^2 = \langle |a_{lm}|^2 \rangle$, where the average is
over all possible observer points.

The scalar amplitude from an inflation model can be calculated analytically
for power-law spectra, giving \cite{LL2}
\begin{equation}
\Sigma_l^2({\rm scalar}) = \frac{\pi}{2} \left[ \frac{\sqrt{\pi}}{2}
	l(l+1) \frac{\Gamma(1+2\epsilon-\eta) \, \Gamma(l-2\epsilon+\eta)}
	{\Gamma(3/2+2\epsilon - \eta) \, \Gamma(l+2+2\epsilon-\eta)} \right]
	\frac{\delta_H^2}{l(l+1)} \,,
\end{equation}
where, for {\em every} multipole, $\delta_H$ is evaluated at the scale $H_0/2$
corresponding to the quad\-rupole. As we are assuming this scale leaves the
horizon 60 $e$-foldings from the end, we have
\begin{equation}
\delta_H^2 = \frac{4}{25\pi} \frac{H_{60}^2}{m_{Pl}^2 \epsilon} \,.
\end{equation}

For the gravitational wave spectrum the general case involves a messy double
integration. To first order in slow-roll we can evade this by using an
approximation due to Lucchin, Matarrese and Mollerach \cite{LMM}, which shows
that if the scalar and tensor power-law indices satisfy $n_{60}^T = n_{60}^S -
1$ (equivalently $\epsilon_{60}'=0$), thus giving power-law inflation, then to
a good approximation the contributions of scalars and tensors to the microwave
multipoles remain in fixed proportion, that proportion being given by
$25\epsilon_{60}/2$. The gravitational wave multipoles can therefore be
generated using the scalar result, but with the spectral index
$1-2\epsilon_{60}$ rather than the true scalar index. The expectations then
add in quadrature to give the total $\Sigma_l^2$.

Finally, one must calculate the prediction for the COBE $10^{\circ}$ result.
The $10^{\circ}$ variance $\sigma_{10}^2$ is given by a weighted sum over the
multipole expectations, where the weighting function $F_l$ corresponds to the
beam profile of the experiment. That is, one writes
\begin{equation}
\sigma_{10}^2 = \frac{1}{2\pi} \sum_l (2l+1) \Sigma_l^2 F_l \,,
\end{equation}
and the COBE weight function is
\begin{equation}
F_l = \frac{1}{2} \exp \left( - \left( \frac{l+1/2}{13.5} \right)^2 \right)
	\,.
\end{equation}

The procedure is now clear. A given model makes a prediction for
$\epsilon_{60}$ and $\eta_{60}$. In all the above expressions, we can pull out
the dependence on $H_{60}$ as a prefactor, and so obtaining the correct
normalization determines $H_{60}$ as a function of $\epsilon_{60}$ and
$\eta_{60}$, and hence directly gives the energy scale at that stage of
inflation.

Throughout we quote figures based on the COBE result $\sigma_{10} = 1.1 \times
10^{-5}$ \cite{COBE}. One has that $H_{60} \propto \sigma_{10}$, and so if
this result is revised one can just scale the results; $H_{60} \rightarrow
H_{60} (\sigma_{10}/1.1 \times 10^{-5})$ and $V_{60}^{1/4} \rightarrow
V_{60}^{1/4} \sqrt{(\sigma_{10}/1.1 \times 10^{-5})}$. This simple scaling
also applies to the values at the end of inflation. If one is merely
interested in upper limits, then one chooses one's preferred upper limit on
$\sigma_{10}$; at present one might advocate the COBE $2$-sigma upper limit of
$1.5 \times 10^{-5}$, and thus say that the inflation scales $V^{1/4}$ are
constrained to be no more than $\sqrt{15/11}$ of the values quoted here.

One should also remember the possibility of cosmic variance --- that a single
observer may see a different $10^{\circ}$ variance than the ensemble average
as calculated above. For the COBE observation, the cosmic variance introduces
an uncertainty of about 10\% \cite{LL2} (weakly dependent on the spectral
indices) which is negligible as compared to the present observational errors
when added in quadrature.

It is unfortunate, but perhaps unsurprising, that the largest values of
$H_{60}$ correspond to large departures from the slow-roll regime, and hence
stretch the validity of approximations used. (We shall see later though that
this is considerably less of a problem with $H_{{\rm end}}$.) It is therefore
necessary in bounding $H_{60}$ to impose some constraints of physical
reasonableness. The choice made here is to assume that the scalar spectral
index lies between about 0.5 and 1.5, as indicated at the $1$-sigma level by
COBE \cite{COBE} (though cases with large gravitational wave contributions
will invalidate their analysis on the scalar index). This therefore requires
$-0.5 \leq 4\epsilon_{60} - 2 \eta_{60} \leq 0.5$. Note that although this is
in principle only a $1$-sigma bound, all inflation-based models with any
chance of satisfying large scale structure data are well within this band
\cite{LL2}, so in fact the limits derived here are almost certainly rather
conservative.

We also impose the additional restriction that $\epsilon_{60}$ and
$|\eta_{60}|$ do not exceed 0.25. The largest values of $H_{60}$ do occur
outside this regime, but some limit must be imposed to assure that the
approximations used do not lose their validity. As we shall see, large values
of the slow-roll parameters are normally not compatible with the requirement
that there be 60 $e$-foldings of inflation to follow, and also these
restrictions are not important in bounding $H_{{\rm end}}$.

The general trends are illustrated in figure 1. In particular one notices the
following properties.
\begin{enumerate}
\item If one fixes a small $\epsilon_{60}$, to exclude gravitational waves,
and varies the tilt using $\eta_{60}$, then one finds a larger $\delta_H$
required as $\eta_{60}$ is made negative, tilting the spectrum to remove short
scale power. The effect on $H_{60}$ is rather modest, however.
\item At fixed tilt ($2\epsilon_{60}-\eta_{60} = {\rm const}$), $\delta_H$ of
course gets smaller as $\epsilon_{60}$ is increased introducing gravitational
waves. However, in determining $H_{60}$ the increasing $\epsilon_{60}$ is a
more important effect (recall $H_{60} \propto \delta_H \sqrt{\epsilon_{60}}$)
and the energy scale $H_{60}$ is increased as $\epsilon_{60}$ and $\eta_{60}$
are increased in concert.
\item The standard normalization $\delta_H = 1.7 \times 10^{-5}$ is accurately
achieved only in a small region about $\epsilon_{60}, |\eta_{60}| \simeq 0$.
Increasing $\epsilon_{60}$ at fixed $\eta_{60}$ decreases it, as does
increasing $\eta_{60}$ at fixed $\epsilon_{60}$.
\end{enumerate}

In the light of this, the largest values of $H_{60}$ come from choosing the
largest $\epsilon_{60}$ and $\eta_{60}$ consistent with the assumptions being
made (trying a large negative $\eta_{60}$ falls foul of the tilt bound). The
maximum value found is $H_{60} = 2.9 \times 10^{-5} m_{Pl}$, corresponding to
a potential energy at that time of $V_{60}^{1/4} = 3.8 \times 10^{16} {\rm
GeV}$, for the values $\epsilon_{60} = \eta_{60} =0.25$.

\section{From $H_{60}$ to $H_{{\rm end}}$}

The COBE normalization gives us specific information about $H_{60}$, dependent
only on $\epsilon_{60}$ and its derivative. To limit the energy at the end of
inflation requires one to evolve the system to $\phi_{{\rm end}}$. At this
point, we remind the reader that inflation can end in two distinct ways
\begin{enumerate}
\item In most slow-rolling models, inflation ends because $\epsilon(\phi)$
grows to equal unity.
\item In certain models such as power-law \cite{LM} and intermediate \cite{BL}
inflation, $\epsilon(\phi)$ never reaches unity in the basic models,
threatening eternal inflation. One escape route often postulated is that the
form of the potential is modified to allow $\epsilon(\phi)$ to increase, which
brings us back to case 1. However, an alternative is that a new mechanism
intervenes to end inflation. The key example is extended inflation \cite{EI},
which looks like power-law inflation in the Einstein conformal frame, but is
brought to an end by the tunnelling of another field with $\epsilon(\phi)$
still small.
\end{enumerate}
We shall largely be concerned with the first, more common case. However, the
results are typically also applicable in the second, as noted below.

The key constraint is that $60$ $e$-foldings remain, which means that
$\epsilon(\phi)$ must satisfy
\begin{equation}
\label{EFOLD}
\frac{60}{\sqrt{4\pi}} = \int_{\phi_{60}}^{\phi_{{\rm end}}}
	\frac{1}{\sqrt{\epsilon(\phi)}} \, \frac{{\rm d}\phi}{m_{Pl}} \,.
\end{equation}
At the same time, we can write
\begin{equation}
\label{HRAT}
\frac{H_{{\rm end}}}{H_{60}} = \exp \left( - \sqrt{4\pi}
	\int_{\phi_{60}}^{\phi_{{\rm end}}} \sqrt{\epsilon(\phi)} \,
	\frac{{\rm d}\phi}{m_{Pl}} \right) \,,
\end{equation}
where $\epsilon(\phi_{{\rm end}}) = 1$ and $H_{60}$ is determined from the
COBE normalization for the given $\epsilon_{60}$ and $\eta_{60}$.

This is most conveniently represented graphically, as in Figure 2, by plotting
$1/\sqrt{\epsilon(\phi)}$ against $\phi/m_{Pl}$. Eq.~(\ref{EFOLD}) then gives
the required area under the curve between the initial value and
$\epsilon(\phi)$ reaching unity. The area under the curve of
$\sqrt{\epsilon(\phi)}$ subject to this constraint measures the reduction of
$H_{{\rm end}}$ relative to $H_{60}$, so one's aim is to minimize this
reduction.

We can see that again constraints of physical reasonableness must be applied
in order to gain worthwhile results. This is because one can always choose
$\epsilon(\phi)$ so as to bring $H_{{\rm end}}$ close to $H_{60}$. This is
done as follows; very rapidly {\em decrease} $\epsilon(\phi)$ until it is
arbitrarily close to zero, keep it there until 60 $e$-foldings have passed,
and then immediately increase it to unity. Of course, this choice is dubious
on physical grounds, and will also violate the initial assumption that the
spectra are power-laws on which the calculation of $H_{60}$ was based, so this
should only be taken as illustrating a general point that by sufficient
contrivance $H_{{\rm end}}$ can always be placed near $H_{60}$.

Let us therefore impose constraints intended to be physically `reasonable' on
the form of $\epsilon(\phi)$. The motivation here lies in assuming that the
form is functionally simple, motivated by the notion that were it not, then
the inferred potential would also be functionally complex, undermining one's
prejudice that it is the potential which belongs to a simple underlying
theory.

\subsubsection*{Case A: $\epsilon(\phi)$ is monotonic}

This follows a suggestion by Lyth \cite{LEPS}, though he employed a slow-roll
approximation. As $\epsilon$ must ultimately increase to unity, and given that
the last 60 $e$-foldings sample only a limited part of the overall potential,
this appears physically well motivated\footnote{It is violated weakly by some
models based on two fields \cite{HYBRID,LL2}, though in any case they tend to
give very low energy scales. The only known example where it is violated
strongly is intermediate inflation \cite{BL}, which can be implemented in an
extended inflation framework and features $\epsilon(\phi)$ decreasing as
$\phi^2$ until tunnelling brings inflation to an end.}. Without paying too
much attention at this point to $\epsilon_{60}'$ beyond noting that $\epsilon'
> 0 \Leftrightarrow \epsilon > \eta$, we can see that the largest final energy
density will be generated if one keeps $\epsilon(\phi)$ at the constant value
$\epsilon_{60}$ until 60 $e$-foldings pass, and then as before increase it
suddenly up to unity. Note that although this would again require an unusual
potential in the single field case, featuring a very flat plateau followed by
a sharp drop, this is in fact exactly what happens in models based on two
fields \cite{HYBRID,LL2}, where inflation driven by the first field ends when
the second field becomes dynamically unstable. This scenario should therefore
certainly not be considered unreasonable.

With these assumptions, it is easy to show that
\begin{equation}
\frac{H_{{\rm end}}}{H_{60}} = \exp \left( - 60 \epsilon_{60} \right) \,.
\end{equation}
This is a very interesting result, because we recall that it was large values
of $\epsilon_{60}$ which gave the largest $H_{60}$, but now we see that such
large values have a detrimental effect on the size of $H_{{\rm end}}$. In
fact, as far as large $H_{{\rm end}}$ is concerned one needs small
$\epsilon_{60}$. The largest $H_{{\rm end}}$ we can achieve is $6.0 \times
10^{-6} m_{Pl}$ for $\epsilon_{60} \simeq 0.007$, a significant reduction on
$H_{60}$. Further, this is for $\eta_{60} = -0.25$, which is not really
consistent with our notion of $\epsilon'$ being small. For more realistic
values of $\eta_{60} \simeq 0$, the limit strengthens yet further to $H_{{\rm
end}} < 4.1 \times 10^{-6} m_{Pl}$, with the maximum at $\epsilon_{60} \simeq
0.008$. Indeed, in the small $\epsilon$, $\eta$ limit this is an analytic
result utilizing the fiducial COBE normalization, from the maximization of
\begin{equation}
H_{{\rm end}}^{{\rm max}} \simeq 7.5 \times 10^{-5} \sqrt{\epsilon_{60}}
	\exp \left(-60 \epsilon_{60} \right) m_{Pl} \,,
\end{equation}
where `max' indicates that this is the maximum possible $H_{{\rm end}}$ for a
given $\epsilon_{60}$. For the maximizing $\epsilon_{60}$, $H_{{\rm end}}$ is
within a factor two of $H_{60}$.

Note that because there is no reduction in $H$ during the rapid growth of
$\epsilon(\phi)$ to unity after 60 $e$-foldings have passed, these limits also
hold in the case where inflation ends with $\epsilon(\phi)$ still less than
one through some additional mechanism, again subject only to the assumption
that $\epsilon'(\phi) \ge 0$. It is interesting to note that extended
inflation features exactly a constant $\epsilon(\phi)$, and hence amongst the
models permitted by the monotonicity assumption it minimizes the reduction of
$H$ during the last 60 $e$-foldings for a given $\epsilon_{60}$.

Recall that this is only subject to the constraint of a monotonic
$\epsilon(\phi)$, making no further assumptions as to the form of the
inflationary potential or approximations to the inflationary dynamics, and
represents a dramatic tightening of the constraints. This is also a good point
to note that there is only an extremely weak dependence, contained in the
exponential which is of order unity, on the assumption that there are 60
$e$-foldings between the quadrupole scale leaving the horizon and the end of
inflation. Different reheating mechanisms have the power to shift this number
by say 10, but this has a negligible impact on the conclusions.

The results illustrate a fundamental point; maximizing $H_{60}$ is not in
general the best way to go about maximizing $H_{{\rm end}}$.

\subsubsection*{Case B: $\epsilon(\phi)$ and $\epsilon'(\phi)$ are monotonic}

This assumption poses yet tighter constraints. In accord with it, the slowest
that $\epsilon(\phi)$ can rise is linearly (from its initial conditions at
$\phi_{60}$), and it is easy to see that linear growth gives the maximum
number of $e$-foldings that could occur. By solving the appropriate equations,
we can get an upper limit on the number of $e$-foldings such a linear
extrapolation would give, as
\begin{equation}
N_{{\rm linear}} < \frac{1}{(\epsilon_{60} - \eta_{60}) \sqrt{\epsilon_{60}}}
	\,.
\end{equation}
If $N_{{\rm linear}}$ is less than sixty, then the construction would be
inconsistent; that is, if one were to keep $\epsilon(\phi)$ and its derivative
monotonic then it would be impossible to achieve 60 $e$-foldings before
inflation ends. This imposes a restriction on the values of $\epsilon_{60}$
and $\eta_{60}$ that are allowed within this assumption. Having satisfied
that, then in accord with the above the smallest reduction in $H$ during the
last 60 $e$-foldings is achieved if $\epsilon(\phi)$ behaves linearly until 60
$e$-foldings have passed, and then increases rapidly to unity. A tedious but
straightforward calculation shows that the reduction factor in this case is
\begin{equation}
\frac{H_{{\rm end}}}{H_{60}} = \exp \left\{ -\frac{\epsilon_{60}}{3 \left(
	\epsilon_{60} - \eta_{60} \right)} \left[ \left( 1 + 60 \left(
	\epsilon_{60} - \eta_{60} \right) \right)^3 - 1 \right] \right\} \,,
\end{equation}
which in the limit $\epsilon_{60}' \rightarrow 0$ (equivalent to
$\epsilon_{60} \rightarrow \eta_{60}$) recovers the result of case A.

Case B tightens the constraint from case A by enforcing that $ \eta_{60}$ be
close to $\epsilon_{60}$, in order to minimize the reduction factor. However,
it does not offer significantly stronger limits than the analytic result
mentioned there for small $|\eta_{60}|$, because the reduction factor is the
same if one chooses $\epsilon_{60} = \eta_{60}$, and it happens that the COBE
normalization does not change much for $\epsilon_{60} = 0.008$ if $\eta_{60}$
is increased from zero to equal $\epsilon_{60}$. With $\epsilon_{60} =
\eta_{60} \simeq 0.008$, we get the maximum value consistent with the case B
assumptions; $H_{{\rm end}} < 4.1 \times 10^{-6} m_{Pl}$. However, in more
specific circumstances the case B assumptions do lead to a tightening of the
limits; for instance if $\epsilon_{60} = 0.008$ and $\eta_{60} = 0$, then the
limit advertized in case A is tightened to 75\% of its case A limit.

\section{Discussion}

In conclusion, limits have been provided on the inflationary energy scale both
60 $e$-foldings from the end of inflation and at the end of inflation. As
noted in the previous section, the results have negligible dependence on the
specific choice of 60 for the number of $e$-foldings between the quadrupole
scale leaving the horizon and the end of inflation, so its dependence on the
details of reheating can be ignored.

Throughout these conclusions, numbers are quoted based on the central COBE
normalization for the $10^{\circ}$ variance; to convert to an upper limit, one
multiplies by the factor by which one is willing to let the true $10^{\circ}$
variance go up. At present, we recommend using the $2$-sigma upper limit,
thus multiplying the numbers for $H$ by $15/11$, though it is worth recalling
that for structure formation models based on inflation the intermediate angle
microwave experiments probably leave little room for the true $10^{\circ}$
variance to be above the COBE result at all \cite{DJ}.

When one incorporates tilt and gravitational wave corrections to the COBE
normalization, one finds that the largest values of $H_{60}$ occur in regions
far from the slow-roll limit, where the validity of the calculations is
breaking down. Nevertheless, by imposing physically motivated constraints
based on prejudice regarding structure formation, it is reasonable to say that
the largest value of $H_{60}$ which can generate the central COBE value is
$H_{60} = 2.9 \times 10^{-5} m_{Pl}$.

The Hamilton-Jacobi equations are used to provide an exact analytic treatment
of the translation of limits on $H_{60}$ into limits on $H_{{\rm end}}$. Again
it is possible by sufficient contrivance in the choice of $\epsilon(\phi)$ to
put $H_{{\rm end}}$ close to $H_{60}$. However, by imposing very reasonable
physically motivated constraints the situation changes dramatically. Here
the properties of the maximization are much nicer, for the maximum values of
$H_{{\rm end}}$ occur in situations where slow-roll was accurately obeyed $60$
$e$-foldings from the end. This fits in with the picture that if slow-roll is
not accurate, then the expansion is far from de Sitter and hence the energy
scale must be decreasing rapidly. The best motivated assumption is that
$\epsilon(\phi)$ monotonically increases with scale, which in general leads to
a maximum $H_{{\rm end}}$ of $6.0 \times 10^{-6} m_{Pl}$. With further
reasonable assumptions this is tightened further to a maximum $H_{{\rm end}}$
of $4.1 \times 10^{-6} m_{Pl}$.

Let us compare these rather abstractly generated limits with the sorts of
values arising in polynomial chaotic inflation models, taking as illustration
$V(\phi) \propto \phi^2$, which coincidentally gives values of $\epsilon_{60}$
and $\eta_{60}$ very similar to those we have advocated as helping to maximize
$H_{{\rm end}}$, though the general $\epsilon(\phi)$ behaviour is of course
different. In section 2, we provided an exact inflationary solution based on
choosing a polynomial $H(\phi)$, which yielded $H_{60} = 6.8 \times 10^{-6}
m_{Pl}$ and $H_{{\rm end}} = 6.2 \times 10^{-7} m_{Pl}$. This solution is a
good approximation to that of a quadratic potential whenever the slow-roll
parameters are small, so the estimate of $H_{60}$ is a good approximation to
that appropriate to $V(\phi) \propto \phi^2$. As slow-roll is a poor
approximation at the end of inflation, the value for $H_{{\rm end}}$ is not as
accurate. Using the normalization at $H_{60}$, but using exact numerical
simulation to evolve to $H_{{\rm end}}$ with the polynomial potential yields
$H_{{\rm end}} = 5.4 \times 10^{-7} m_{Pl}$.

Throughout, we have been providing what amounts to upper limits on the energy
scale, by finding the largest values of the energy scale consistent with the
COBE normalization and various dynamical constraints. No mention has yet been
made of lower limits, for the reason that the energy scale can be made as low
as one likes while still satisfying COBE, provided one is willing to accept
very small values of $\epsilon_{60}$. As models do exist where $\epsilon_{60}$
can be tiny (such a class are the `hybrid' models featuring one inflaton field
and a trigger field to end inflation \cite{HYBRID}, and natural inflation
\cite{NAT} provides a further example), lower limits cannot be derived using
COBE alone. However, there is one very promising route by which a lower limit
could be placed, which would be if it were to be demonstrated that some
sizeable component of the COBE result were due to gravitational waves
\cite{CBDES}. Such a discovery effectively places a lower limit on
$\epsilon_{60}$, and hence on the inflationary energy scale. It has already
been noted that some knowledge of tensor modes is essential if one hopes to
determine the detailed form of the inflaton potential \cite{REC}.

Limits on $\rho_{{\rm end}}$ can be converted to limits on the reheat
temperature $T_{{\rm reh}}$, given two uncertainties. The first is that the
energy is to be distributed evenly amongst some unknown number $g_*$ of
particle degrees of freedom available at that energy; $g_*$ is assumed to be
at least the standard model value of 106.75 but could be much larger.
Secondly, there is a parameter $\alpha < 1$ which measures the efficiency of
reheating, $\rho_{{\rm reh}} = \alpha \rho_{{\rm end}}$, where $\rho_{{\rm
reh}}$ is the energy density when the post-inflationary thermalization can
first be said to have completed. In weakly coupled theories $\alpha$ is
expected to be rather small, though in theories where inflation ends
violently, such as through bubble collisions, it may not be too far from
unity. Putting all this together gives
\begin{equation}
\frac{T_{{\rm reh}}}{m_{Pl}} \simeq 0.78 g_*^{-1/4} \alpha^{1/4} \left(
	\frac{H_{{\rm end}}}{m_{Pl}} \right)^{1/2} \,.
\end{equation}
Making the weak, but not essential, assumption that $\epsilon'(\phi) \ge 0$
during the last 60 $e$-foldings of inflation, and using the standard model
degrees of freedom, it is reasonable to expect that the reheat temperature
after inflation will not exceed
\begin{equation}
T_{{\rm reh}} = 7.2 \alpha^{1/4} \times 10^{15} {\rm GeV} \,.
\end{equation}

We end with some brief comments concerning topological defects. It has been
shown that typically one needs a defect scale of slightly over $10^{16}$ GeV
if defects are to explain large scale structure \cite{DEFENG}. It is clear
that forming such defects after reheating will be very difficult here, because
as a first step one must reduce the inflationary contribution to COBE to
almost negligible size (as topological defect theories are already likely to
produce excessive distortions if normalized to other large scale structure
data). The reheat temperature comes down as the square root of the fractional
lowering of the COBE signal, so to remove the inflationary density
perturbations will bring down the reheat temperature by another factor of at
least 3. There is however another possibility which can be realised with
particular ease in hybrid models \cite{HYBRID}, which is to form defects as
inflation ends in the field which is triggering the end of inflation. In that
case typically all the inflationary energy density is available to go into
defects, evading both the $g_*$ and $\alpha$ suppression factors, and removing
the need to restore the symmetry. Given the tight constraints illustrated
above, this seems the most promising route to salvaging compatability of
defect theories with the inflationary cosmology.

\section*{Acknowledgements}
The author is supported by the SERC, and acknowledges the use of the
Starlink computer system at the University of Sussex. I am especially grateful
to Jim Lidsey and David Lyth for discussions relating to several aspects of
this work. I would also like to thank Jaume Garriga for discussions which
prompted me to look at this problem in the first place, and David Wands for a
careful reading of the manuscript.
\frenchspacing

\newpage
\nonfrenchspacing
\section*{Figure Captions}

\noindent
{\em Figure 1a,b}

These figures give the COBE normalization of $H_{60}$ (solid lines) and the
corresponding scalar spectrum normalization $\delta_H$ (dashed lines), as a
function of $\epsilon_{60}$ and $\eta_{60}$ (the numerical values being
conveniently close to each other). These include both tilted spectrum and
gravitational wave corrections, as described in text. Figure 1a plots them
verses $\epsilon_{60}$ for various fixed $\eta_{60}$, and Figure 1b verses
$\eta_{60}$ at various fixed $\epsilon_{60}$. The trends are summarized in the
text.

\vspace{0.5cm}
\noindent
{\em Figure 2}

A graphical illustration of Eqs.~(\ref{EFOLD}) and (\ref{HRAT}). The value of
$\phi$ can be shifted by the addition of an arbitrary constant. At $\phi_{60}$
one has some particular values for $\epsilon_{60}$ and $\epsilon_{60}'$. As
$\phi$ increases, $1/\sqrt{\epsilon(\phi)}$ must vary in such a way that the
area under it reaches $60/\sqrt{4\pi}$ just as $\epsilon(\phi)$ reaches unity
to end inflation (in general it need not do so monotonically as illustrated
here). At the same time, the area under the curve $\sqrt{\epsilon(\phi)}$
measures the decrease of $H$ relative to $H_{60}$ in accord with
Eq.~(\ref{HRAT}). It is clear that the smallest decrease in $H$ during the
last 60 $e$-foldings is achieved by keeping $\epsilon(\phi)$ as small as
possible for as long as possible.

\end{document}